\documentclass[10pt,letterpaper]{iopart}
\usepackage{setstack, iopams, graphicx, hyperref}
\usepackage[caption=false]{subfig}
\begin{document}

%-------------------------------------------------------------------------------
% Shortcut Commands
%-------------------------------------------------------------------------------

\newcommand{\braket}[2]{{\left\langle #1 \middle| #2 \right\rangle}}
\newcommand{\bra}[1]{{\left\langle #1 \right|}}
\newcommand{\ket}[1]{{\left| #1 \right\rangle}}
\newcommand{\ketbra}[2]{{\left| #1 \middle\rangle \middle \langle #2 \right|}}

%-------------------------------------------------------------------------------
% Front Matter
%-------------------------------------------------------------------------------

\title{Coined Quantum Walks on Weighted Graphs}

\author{Thomas G Wong}
\address{Department of Computer Science, University of Texas at Austin, 2317 Speedway, Austin, Texas 78712, USA}
\address{Currently at Department of Physics, Creighton University, 2500 California Plaza, Omaha, Nebraska 68178, USA}
\ead{\mailto{thomaswong@creighton.edu}}

\begin{abstract}
	We define a discrete-time, coined quantum walk on weighted graphs that is inspired by Szegedy's quantum walk. Using this, we prove that many lackadaisical quantum walks, where each vertex has $l$ integer self-loops, can be generalized to a quantum walk where each vertex has a single self-loop of real-valued weight $l$. We apply this real-valued lackadaisical quantum walk to two problems. First, we analyze it on the line or one-dimensional lattice, showing that it is exactly equivalent to a continuous deformation of the three-state Grover walk with faster ballistic dispersion. Second, we generalize Grover's algorithm, or search on the complete graph, to have a weighted self-loop at each vertex, yielding an improved success probability when $l < 3 + 2\sqrt{2} \approx 5.828$.
\end{abstract}

\pacs{03.67.Ac, 02.10.Ox}

%-------------------------------------------------------------------------------
% Main Matter
%-------------------------------------------------------------------------------

\section{Introduction}

Quantum walks are a universal model of quantum computation \cite{Childs2009} that have been used to develop a variety of quantum algorithms, including for searching \cite{SKW2003}, element distinctness \cite{Ambainis2004}, and boolean formula evaluation \cite{FGG2008}. In these algorithms, the quantum walk occurs on unweighted graphs, or equivalently where each edge of the graph has unit weight.

Despite this success of quantum walks on unweighted graphs, several papers have identified improvements for quantum walks when using weighted graphs. For continuous-time quantum walks, this includes state transfer \cite{Christandl2004,Kendon2011}, universal mixing \cite{Carlson2007}, and searching \cite{Wong16,Wong22}. For discrete-time quantum walks, Szegedy's quantum walk \cite{Szegedy2004} is naturally defined on weighted graphs by quantizing Markov chains with arbitrary transition amplitudes. Although Szegedy's quantum walk is equivalent to a coined quantum walk under certain conditions (\cite{Magniez2012} in the context of searching, and \cite{Wong26} for undirected and unweighted Markov chains), this equivalence has received less attention than it deserves. As a result, coined quantum walks are usually not understood, investigated, or interpreted as walks on weighted graphs. In this paper, we remedy this by giving an explicit definition of a coined quantum walk on a weighted graph that is inspired by Szegedy's quantum walk. When this coined quantum walk on a weighted graph uses the flip-flop shift to hop between vertices, it is exactly equivalent to Szegedy's quantum walk. When it uses the moving shift, however, it differs and is a new type of walk. In the next section, we discuss this in detail by reviewing the definition of the coined quantum walk, Szegedy's quantum walk, and their precise equivalence \cite{Wong26}. Then we define the coined quantum walk on weighted graphs by taking inspiration from Szegedy's quantum walk. Then in section 3, we analyze the coin operator for the coined quantum walk on weighted graphs, showing that it no longer performs an inversion about the average that the unweighted coin does. 

In section 4, we prove that if multiple edges of an unweighted graph evolve identically, they can be replaced with a single weighted edge. This naturally leads to a generalization of the lackadaisical quantum walk \cite{Wong10}. The lackadaisical quantum walk is a quantum analogue of a lazy random walk, where each vertex is given $l$ integer self-loops, so the particle has some amplitude of staying put. Typically, the $l$ self-loops at each vertex of a lackadaisical quantum walk evolve identically, so they can be replaced by a single self-loop of weight $l$ at each vertex. Then when $l$ is an integer, it reproduces the behavior of the original lackadaisical quantum walk. But now $l$ can also take non-integer real values, thus defining a generalization of the lackadaisical quantum walk to real-valued weights.

We analyze this generalized lackadaisical quantum walk for two problems. In section 5, we consider the walk on the line or one-dimensional (1D) lattice. For the standard lackadaisical quantum walk with an integer number of self-loops per vertex, this was investigated by \cite{Wang2016}, who showed that the ballistic dispersion can be faster than, slower than, or the same as the loopless walk. Replacing the $l$ integer self-loops with a single self-loop of weight $l$, we show that the generalized lackadaisical quantum walk on the line is exactly equivalent to a continuous deformation of the three-state Grover walk proposed by {\v{S}}tefa{\v{n}}{\'a}k \textit{et al} \cite{Stefanak2012}.

In section 6, we investigate quantum search on the complete graph, which is the quantum walk-formulation of Grover's unstructured search algorithm \cite{Grover1996}. This was the problem that introduced lackadaisical quantum walks \cite{Wong10}, and with $l$ integer self-loops per vertex, the success probability is improved over the loopless value when $l \le 5$ \cite{Wong10}. Generalizing the lackadaisical quantum walk so that $l$ can take non-integer values, we determine the runtime and success probability of the algorithm for all real $l \ge 0$, and this includes qualitatively new behavior when $l < 1/3$. This reveals that the success probability is improved over the loopless case when $l < 3 + 2\sqrt{2} \approx 5.828$. Finally, we end in section 7 with concluding remarks.

%-------------------------------------------------------------------------------
% Section
%-------------------------------------------------------------------------------

\section{Definition}

The coined quantum walk was first introduced by Meyer \cite{Meyer1996a} in the context of quantum cellular automata, who showed that an internal degree of freedom allowed the particle to evolve nontrivially. This was later framed in the language of quantum walks by Aharonov \textit{et al}~\cite{Aharonov2001}. The particle hops on the $N$ vertices of a graph, and the internal coin state identifies the directions in which the particle can hop. We write the states as $\ket{v} \otimes \ket{u} = \ket{vu}$ to denote a particle at vertex $v$ pointing towards vertex $u$. The quantum walk is effected by a coin flip $C$ and a shift $S$, so one step of the walk is
\begin{equation}
	\label{eq:U}
	U = SC.
\end{equation}
The coin operator $C$ applies a coin $C_v$ to each vertex $v$:
\begin{equation}
	\label{eq:C}
	C  = \sum_v \ketbra{v}{v} \otimes C_v,
\end{equation}
where $C_v$ is typically the Grover diffusion coin \cite{Moore2002}
\begin{equation}
	\label{eq:Cv}
	C_v = 2 \ketbra{s_v}{s_v} - I
\end{equation}
that reflects across
\begin{equation}
	\label{eq:sv_unweighted}
	\ket{s_v} = \frac{1}{\sqrt{\deg(v)}} \sum_{u \sim v} \ket{u} \quad {\rm (unweighted).}
\end{equation}
This corresponds to a quantum walk on an unweighted graph because $\ket{s_v}$ is the uniform, unweighted superposition of directions at $v$.

For the shift $S$, two different operators are often used. The first is the ``moving shift'' on a lattice, where a particle hops and keeps pointing in the same direction \cite{Meyer1996a,Ambainis2001}. For example, on the line, a particle pointing right will hop to the right and continue pointing right, so $S\ket{0,1} = \ket{1,2}$. The second is the ``flip-flop shift,'' where the particle hops and turns around \cite{AKR2005}, so $S\ket{0,1} = \ket{1,0}$. The flip-flop shift is more straightforward than the moving shift on nonlattice graphs, and it is needed for fast quantum search on lattices \cite{AKR2005}.

To define a coined quantum walk on weighted graphs, we will need to generalize $\ket{s_v}$ \eref{eq:sv_unweighted} to weighted graphs, which in turn changes the coin operator \eref{eq:Cv}. The shift does not need to be changed. We can determine $\ket{s_v}$ for a weighted graph using Szegedy's quantum walk.

\begin{figure}
\begin{center}
	\subfloat[]{
		\includegraphics{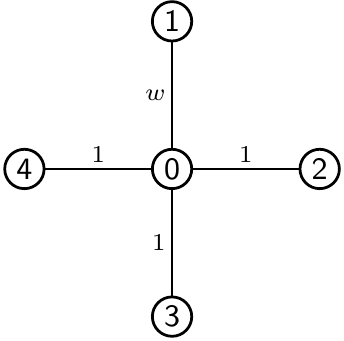}
		\label{fig:graph}
	} \quad \quad \quad
	\subfloat[]{
		\includegraphics{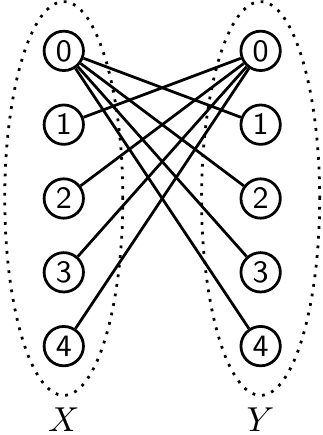}
		\label{fig:graph_szegedy}
	}
	\caption{\label{fig:graph_together} (a) A weighted graph and (b) its bipartite double cover.}
\end{center}
\end{figure}

Szegedy's quantum walk \cite{Szegedy2004} is a quantization of a classical random walk or Markov chain. For example, consider the weighted graph in \fref{fig:graph}. Its classical transition probabilities are
\begin{eqnarray*}
	P_{01} = \frac{w}{w+3}, \quad P_{02} = P_{03} = P_{04} = \frac{1}{w+3}, \\
	P_{10} = P_{20} = P_{30} = P_{40} = 1,
\end{eqnarray*}
where $P_{ij}$ is the probability of transitioning from vertex $i$ to $j$, and the remaining transitions have probability 0. Szegedy's quantum walk occurs on the edges of the bipartite double cover of the original graph, which is shown in \fref{fig:graph_szegedy}. If we label the partite sets $X$ and $Y$, the walk evolves by repeated applications of
\[ W = R_2 R_1, \]
where the operators
\begin{eqnarray}
	R_1 = 2 \sum_{x \in X} \ketbra{\phi_x}{\phi_x} - I, \label{eq:R1} \\
	R_2 = 2 \sum_{y \in Y} \ketbra{\psi_y}{\psi_y} - I, \nonumber
\end{eqnarray}
are reflections across the states
\begin{eqnarray*}
	\ket{\phi_x} = \ket{x} \otimes \sum_{y \in Y} \sqrt{P_{xy}} \ket{y}, \\
	\ket{\psi_y} = \sum_{x \in X} \sqrt{P_{yx}} \ket{x} \otimes \ket{y},
\end{eqnarray*}
where $\ket{x} \otimes \ket{y} = \ket{xy}$ denotes the edge connecting vertices $x \in X$ and $y \in Y$. For example, for the weighted graph in \fref{fig:graph_together}, $R_1$ with $x = 0$ reflects across the state
\[ \ket{\phi_0} = \ket{0} \otimes \frac{1}{\sqrt{w+3}} \left( \sqrt{w} \ket{1} + \ket{2} + \ket{3} + \ket{4} \right). \]

For unweighted graphs, it is known that two steps of the coined quantum walk with the flip-flop shift are equivalent to one step of Szegedy's quantum walk \cite{Magniez2012,Wong26}. More precisely, using the bijection that an edge connecting $x \in X$ to $y \in Y$ (in the bipartite double cover) in Szegedy's quantum walk corresponds to a coined particle at vertex $x$ pointing to vertex $y$ (in the original graph), the relationships between the operators are $C = R_1$ and $SCS = R_2$ \cite{Wong26}. Then $U^2 = (SC)(SC) = (SCS)C = R_2 R_1 = W$, so two steps of the coined quantum walk are equal to one step of Szegedy's.

To preserve this relationship for weighted graphs, we use $C = R_1$ to equate \eref{eq:C} and \eref{eq:R1}. Doing this, we identify
\begin{equation}
	\label{eq:sv}
	\ket{s_v} = \sum_{u \sim v} \sqrt{P_{vu}} \ket{u} = \frac{1}{\sqrt{\sum_t w_{vt}}} \sum_{u \sim v} \sqrt{w_{vu}} \ket{u} \quad {\rm (weighted)},
\end{equation}
where $w_{vu}$ is the weight of the edge connecting vertices $v$ and $u$. For example, for the weighted graph in \fref{fig:graph},
\[ \ket{s_0} = \frac{1}{\sqrt{w+3}} \left( \sqrt{w} \ket{1} + \ket{2} + \ket{3} + \ket{4} \right). \]
When the graph is unweighted (i.e., $w_{vu} = 1$ for all $v \sim u$), \eref{eq:sv} reduces to the uniform superposition \eref{eq:sv_unweighted}. This yields our definition of a coined quantum walk on a weighted graph: We evolve by repeatedly applying \eref{eq:U} with the Grover diffusion coin $C_v$ \eref{eq:Cv} that reflects across the weighted state \eref{eq:sv}.

This definition of a coined quantum walk on a weighted graph preserves the relationships $C = R_1$ and $SCS = R_2$ when $S$ is the flip-flop shift, so two steps of the coined quantum walk with the flip-flop shift are equal to one step of Szegedy's, even for weighted graphs. If the moving shift is used, however, it is a different walk.

%-------------------------------------------------------------------------------
% Section
%-------------------------------------------------------------------------------

\section{Weighted Coin and Inversions}

Now let us examine in detail how the weighted coin operator reflects across \eref{eq:sv} and how it differs from the unweighted coin. As a concrete example, let us again consider the graph in \fref{fig:graph}, and say the particle is at vertex $0$ and points to vertices $1$, $2$, $3$, and $4$ with respective amplitudes $\alpha_1$, $\alpha_2$, $\alpha_3$, and $\alpha_4$. That is, the particle is in the state $\ket{\psi} = \ket{0} \otimes \ket{\psi_0}$ with
\[ \ket{\psi_0} =  \alpha_1 \ket{1} + \alpha_2 \ket{2} + \alpha_3 \ket{3} + \alpha_4 \ket{4}. \]
The weighted coin acts on this state by
\begin{eqnarray*}
	C \ket{\psi} 
		&= \ket{0} \otimes C_0 \ket{\psi_0} \\
		&= \ket{0} \otimes \left( 2 \ket{s_0} \braket{s_0}{\psi_0} - \ket{\psi_0} \right) \\
		&= \ket{0} \otimes \left( 2 \ket{s_0} \frac{\sqrt{w} \alpha_1 + \alpha_2 + \alpha_3 + \alpha_4}{\sqrt{w+3}} - \ket{\psi_0} \right) \\
		&= \ket{0} \otimes \left( 2 \bar{\alpha} \sqrt{w+3} \ket{s_0} - \ket{\psi_0} \right) \\
		&= \ket{0} \otimes \Big[ (2\bar{\alpha}\sqrt{w} - \alpha_1) \ket{1} + (2\bar{\alpha} - \alpha_2) \ket{2} + (2\bar{\alpha} - \alpha_3) \ket{3} \\
		&\quad\quad\quad\quad + (2\bar{\alpha} - \alpha_4) \ket{4} \Big],
\end{eqnarray*}
where
\[ \bar{\alpha} = \frac{\sqrt{w} \alpha_1 + \alpha_2 + \alpha_3 + \alpha_4}{w+3}. \]
Since $2\bar{\alpha}\sqrt{w} - \alpha_1$ is an inversion of $\alpha_1$ about $\bar{\alpha}\sqrt{w}$, we see that the amplitude pointing towards $\ket{1}$ is inverted about $\bar{\alpha}\sqrt{w}$, and the other amplitudes are inverted about $\bar{\alpha}$.

Generalizing this to an arbitrary weighted graph, if the state of the system is
\[ \ket{\psi} = \sum_v \ket{v} \otimes \sum_{u \sim v} \alpha_{vu} \ket{u}, \]
then the coin inverts the amplitudes to yield
\begin{equation}
	\label{eq:invert}
	C \ket{\psi} = \sum_v \ket{v} \otimes \sum_{u \sim v} ( 2\bar\alpha_v \sqrt{w_{vu}} - \alpha_{vu} ) \ket{u},
\end{equation}
where
\begin{equation}
	\label{eq:bar}
	\bar\alpha_v = \frac{\sum_{u \sim v} \sqrt{w_{vu}} \alpha_{vu}}{\sum_{u \sim v} w_{vu}}.
\end{equation}

If the graph is unweighted, so $w_{vu} = 1$ for all $u \sim v$, then $\bar{\alpha}_v$ is the average of the amplitudes pointing from $v$ to its neighbors. Then the amplitudes are inverted about their mean (see Lemma 2 of \cite{Wong23}), and this is akin to the ``inversion about the mean'' of Grover's algorithm \cite{Grover1996}. With weights, $\bar{\alpha}_v$ is no longer the mean, so the coin is no longer an inversion about the mean.

%-------------------------------------------------------------------------------
% Section
%-------------------------------------------------------------------------------

\section{\label{sec:reduction} Identically-Evolving Edges to a Weighted Edge}

In this section, we show that some quantum walks on unweighted graphs can be reinterpreted as quantum walks on weighted graphs by combining identically-evolving amplitudes. When we apply this result to lackadaisical quantum walks, it allows us to replace the $l$ self-loops at a vertex with one self-loop of weight $l$.

Say we have an unweighted graph where a vertex $v$ has degree $d$, so a particle at vertex $v$ can point in $d$ different directions. Then the state of a particle at vertex $v$ can be written as
\begin{equation}
	\label{eq:psi_full}
	\ket{\psi} = \ket{v} \otimes \left( \alpha_1 \ket{1} + \alpha_2 \ket{2} + \dots + \alpha_{d} \ket{d} \right).
\end{equation}
The Grover diffusion coin transforms this to
\begin{equation}
	\label{eq:Cpsi_full}
	C \ket{\psi} = \ket{v} \otimes \left[ (2 \bar{\alpha} - \alpha_1) \ket{1} + \dots + (2 \bar{\alpha} - \alpha_d) \ket{d} \right],
\end{equation}
where
\begin{equation}
	\label{eq:bar_full}
	\bar{\alpha} = \frac{\alpha_1 + \alpha_2 + \dots + \alpha_d}{d}
\end{equation}
is the average of the amplitudes \cite{Wong23}. So it inverts each amplitude about the average.

Now say $k$ of the amplitudes evolve identically due to the symmetry of the graph. For example, for a lackadaisical quantum walk, the amplitudes along the $l$ self-loops at a vertex often evolve identically. Without loss of generality, say this corresponds to the first $k$ basis states, i.e.,
\[ \alpha_1 = \alpha_2 = \dots = \alpha_k. \]
Using this, let us reduce \eref{eq:psi_full}, \eref{eq:Cpsi_full}, and \eref{eq:bar_full} to a subspace. The state of the system \eref{eq:psi_full} can be written as
\begin{eqnarray*}
	\ket{\psi} 
		&= \ket{v} \otimes \left[ \alpha_1 ( \ket{1} + \dots + \ket{k} ) + \alpha_{k+1} \ket{k+1} + \dots + \alpha_d \ket{d} \right] \\
		&= \ket{v} \otimes \left[ \alpha_1 \sqrt{k} \ket{\sigma} + \alpha_{k+1} \ket{k+1} + \dots + \alpha_d \ket{d} \right], \\
		&= \ket{v} \otimes \left[ \alpha_\sigma \ket{\sigma} + \alpha_{k+1} \ket{k+1} + \dots + \alpha_d \ket{d} \right],
\end{eqnarray*}
where
\[ \ket{\sigma} = \frac{1}{\sqrt{k}} \left( \ket{1} + \dots + \ket{k} \right) \]
is the uniform superposition of the identically-evolving states, and
\[ \alpha_\sigma = \alpha_1 \sqrt{k} \]
is its corresponding amplitude. So we have written the state $\ket{\psi}$ in a $(d-k+1)$-dimensional subspace. Now for \eref{eq:Cpsi_full}, we can reduce it to the same subspace:
\begin{eqnarray*}
	C \ket{\psi} 
		&= \ket{v} \otimes \Big[ (2\bar{\alpha} - \alpha_1) ( \ket{1} + \dots + \ket{k} ) + (2\bar{\alpha} - \alpha_{k+1}) \ket{k+1} + \dots \\
		&\quad\quad\quad\quad + (2\bar{\alpha} - \alpha_k) \ket{k} \Big] \\
		&= \ket{v} \otimes \Big[ (2\bar{\alpha} - \alpha_1) \sqrt{k} \ket{\sigma} + (2\bar{\alpha} - \alpha_{k+1}) \ket{k+1} + \dots \\
		&\quad\quad\quad\quad + (2\bar{\alpha} - \alpha_k) \ket{k} \Big], \\
		&= \ket{v} \otimes \Big[ (2\bar{\alpha}\sqrt{k} - \alpha_\sigma) \ket{\sigma} + (2\bar{\alpha} - \alpha_{k+1}) \ket{k+1} + \dots \\
		&\quad\quad\quad\quad + (2\bar{\alpha} - \alpha_k) \ket{k} \Big].
\end{eqnarray*}
Finally, we can rewrite \eref{eq:bar_full} as
\[ \bar{\alpha} = \frac{\alpha_1 k + \alpha_{k+1} + \dots + \alpha_d}{d} = \frac{\alpha_\sigma \sqrt{k} + \alpha_{k+1} + \dots + \alpha_d}{d}. \]
Comparing this with the inversion of a weighted graph in \eref{eq:invert} and \eref{eq:bar}, we see that the uniform superposition of identically-evolving edges $\ket{\sigma}$ evolves exactly like an edge of weight $k$, while the remaining edges continue to be unweighted. Thus $k$ identically-evolving, unweighted edges can be replaced by a single edge of weight $k$. Hence for a lackadaisical quantum walk, if the $l$ self-loops at a vertex evolve identically, they can be replaced by a single self-loop of weight $l$. This generalizes the values that $l$ can take from the integers to the reals.

In the next two sections, we apply this reduction and generalization of the lackadaisical quantum walk to two examples. The first is the quantum walk on the line, whose generalization is exactly equivalent to another type of quantum walk. The second is quantum search on the complete graph, where additional analysis is needed to get the runtime and success probability for all values of $l$.

%-------------------------------------------------------------------------------
% Section
%-------------------------------------------------------------------------------

\section{Walk on the Line}

\begin{figure}
\begin{center}
	\subfloat[]{
		\includegraphics{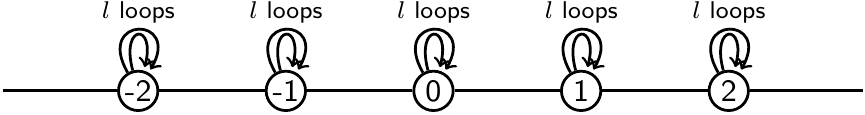}
		\label{fig:line_loops}
	}

	\subfloat[]{
		\includegraphics{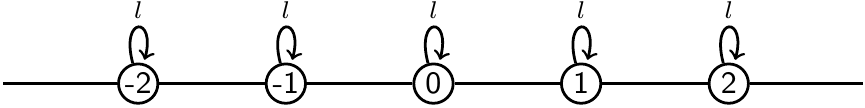}
		\label{fig:line_loop}
	}
	\caption{A one-dimensional line or lattice with (a) $l$ self-loops per vertex and (b) a self-loop of weight $l$ at each vertex.}
\end{center}
\end{figure}

\begin{figure}
\begin{center}
	\includegraphics{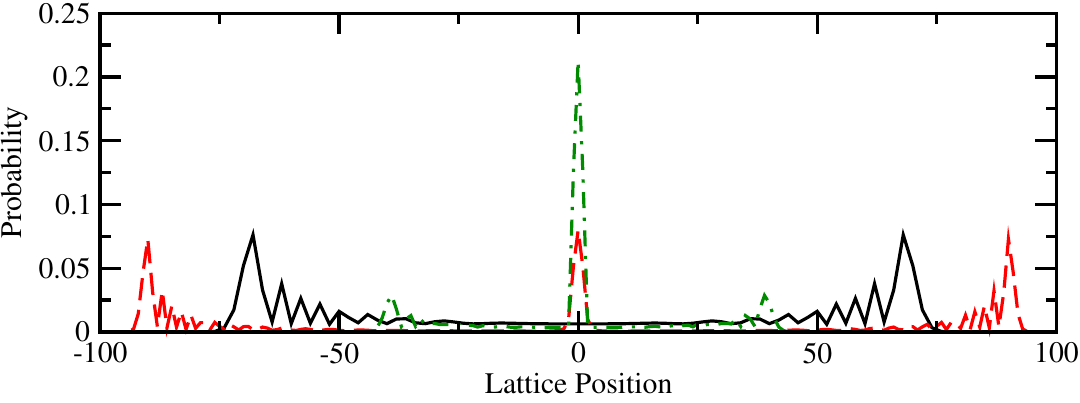}
	\caption{\label{fig:line_T100} Probability distribution for a quantum walk on the line after 100 steps with the initial state $\ket{0} \otimes \left( \ket{-1} + i\ket{1} \right) / \sqrt{2}$. The solid black curve is without self-loops and only includes even locations since the probability at odd locations is zero. The dashed red curve is a lackadaisical quantum walk with $l = 10$ with the moving shift, and the dot-dashed green curve is with the flip-flop shift.}
\end{center}
\end{figure}

For our first example, consider a quantum walk on the line. The lackadaisical case of $l$ self-loops per vertex, as depicted in \fref{fig:line_loops}, was explored by \cite{Wang2016} with the moving shift. Their initial state was a particle localized at vertex $0$ with amplitude in the left- and right-moving coin states only, so there was no initial amplitude along the self-loops. They showed that the more self loops, the faster the ballistic dispersion, with the velocities of the peaks equal to $\pm \sqrt{l/(l+2)}$. This is illustrated in \fref{fig:line_T100}, where the dashed red curve with $l = 10$ spreads more quickly than the loopless solid black curve.

Due to the symmetry of the evolution, the amplitudes along the $l$ self-loops at a vertex evolve identically. Then from \sref{sec:reduction}, we can replace them with a single self-loop of weight $l$, as shown in \fref{fig:line_loop}. The results from \cite{Wang2016} carry over to this generalized case, such as the peak velocities being $\pm \sqrt{l/(l+2)}$, except $l$ can now take non-integer values.

Let us analyze this generalized lackadaisical quantum walk more closely so that we can connect it to other work. At each vertex, a particle can point to the left, to itself through the weighted self-loop (i.e., stay), or to the right. Let us call these coin states $\ket{L}$, $\ket{S}$, and $\ket{R}$. Then $\ket{s_v}$ \eref{eq:sv} for each vertex $v$ is
\[ \ket{s_v} = \frac{1}{\sqrt{l+2}} \left( \ket{L} + \sqrt{l} \ket{S} + \ket{R} \right). \]
Then in the $\{ \ket{L}, \ket{S}, \ket{R} \}$ basis, $C_v$ \eref{eq:Cv} can be written as a $3 \times 3$ matrix
\[ C_v = \frac{1}{l + 2} \left( \begin{array}{ccc}
		-l & 2\sqrt{l} & 2 \\
		2\sqrt{l} & l-2 & 2\sqrt{l} \\
		2 & 2\sqrt{l} & -l \\
\end{array} \right). \]
After applying $C_v$ at each vertex $v$, the moving shift is applied, completing a step of the quantum walk $U = SC$ \eref{eq:U}.

Now let us consider another type of quantum walk called a three-state quantum walk \cite{Inui2005}, which we will later prove is equivalent to the above lackadaisical quantum walk on the line. The three-state quantum walk is a quantum walk on the line with three internal coin states, one pointing left, one to stay, and one pointing right, corresponding to the coin basis states $\ket{L}$, $\ket{S}$, and $\ket{R}$. The original three-state quantum walk \cite{Inui2005} simply applies the unweighted Grover diffusion coin \eref{eq:sv_unweighted} followed by the moving shift. {\v{S}}tefa{\v{n}}{\'a}k \textit{et al}, however, considered the 1D walk with deformations of the Grover coin, firstly by deforming its eigenvalues and secondly by deforming its eigenvectors. The second deformation uses the coin operator given by Eq.~(14) of \cite{Stefanak2012}:
\[ C_v = \left( \begin{array}{ccc}
	-\rho^2 & \rho\sqrt{2(1-\rho^2)} & 1-\rho^2 \\
	\rho\sqrt{2(1-\rho^2)} & 2\rho^2-1 & \rho\sqrt{2(1-\rho^2)} \\
	1-\rho^2 & \rho\sqrt{2(1-\rho^2)} & -\rho^2 \\
\end{array} \right), \]
where $\rho \in [0,1]$ is a continuous parameter with the unweighted Grover diffusion coin corresponding to $\rho = 1/\sqrt{3}$. The coin is followed by the moving shift, and they showed that the velocity at which the peaks travel is $\pm \rho$. So as $\rho$ increases, the speed of the ballistic dispersion also increases.

{\v{S}}tefa{\v{n}}{\'a}k \textit{et al}'s deformed coin operator equals the 1D lackadaisical quantum walk's coin with
\[ \rho = \sqrt{\frac{l}{l+2}}. \]
So their walk is exactly a generalized lackadaisical quantum walk with a weighted self-loop, and all their results carry over. This gives a new interpretation of {\v{S}}tefa{\v{n}}{\'a}k \textit{et al}'s result, not as an eigenvector deformation, but as a coined quantum walk on the line with a weighted self-loop at each vertex.

The improved dispersion of the lackadaisical quantum walk on the line necessitates the moving shift. In the loopless case, both the moving and flip-flop shifts effect the same evolution for the initially unbiased state \cite{Wong17}. With self-loops, however, their evolutions are significantly different, as shown in \fref{fig:line_T100}, where the dashed red curve corresponds to the moving shift and the dot-dashed green curve to the flip-flop shift, both with $l = 10$. Whereas the moving shift's dispersion is faster, the flip-flop shift's is slower.

%-------------------------------------------------------------------------------
% Section
%-------------------------------------------------------------------------------

\section{Search on the Complete Graph}

\begin{figure}
\begin{center}
	\subfloat[]{
		\includegraphics{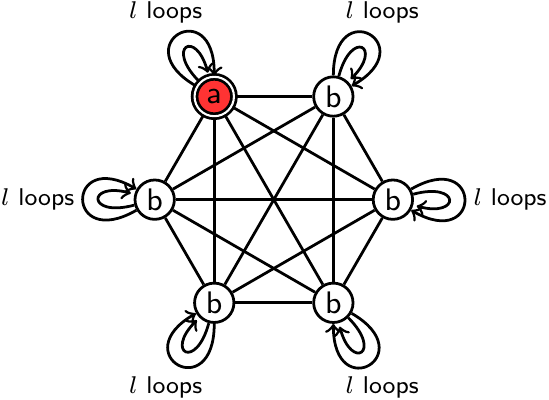}
		\label{fig:complete_loops}
	} \quad \quad
	\subfloat[]{
		\includegraphics{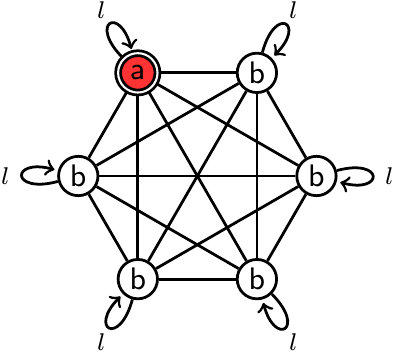}
		\label{fig:complete_loop}
	}
	\caption{The complete graph of $N = 6$ vertices with (a) $l$ self-loops per vertex and (b) a self-loop of weight $l$ at each vertex. A marked vertex is indicated by a double circle. Identically evolving vertices are identically colored and labeled.}
\end{center}
\end{figure}

For our second example, consider search on the complete graph, which corresponds to Grover's unstructured search problem, by lackadaisical quantum walk. This was first explored in \cite{Wong10} with $l$ self-loops per vertex, as illustrated in \fref{fig:complete_loops}. Due to the symmetry of the problem, at each vertex, the amplitudes along the $l$ self-loops evolve identically. Then from \sref{sec:reduction}, we can replace them with a single self-loop of weight $l$, as shown in \fref{fig:complete_loop}. The analysis from \cite{Wong10} carries over to our generalized walk with real $l$ when $l \ge 1/3$. We will give new analysis for $l < 1/3$, thus completely characterizing the runtime and success probability of the algorithm for all real $l \ge 0$.

First let us carry over the results from \cite{Wong10}. As shown there, from the symmetry of the problem, there are only two types of vertices, the marked $a$ vertex and the unmarked $b$ vertices, depicted in \fref{fig:complete_loop}. A particle at the $a$ vertex can either point to itself or to $b$ vertices, and a particle at a $b$ vertex can either point to the $a$ vertex or to $b$ vertices (including itself). Then the system evolves in a 4D subspace spanned by
\begin{eqnarray*}
	\ket{aa} = \ket{a} \otimes \ket{a}, \\
	\ket{ab} = \ket{a} \otimes \frac{1}{\sqrt{N-1}} \sum_{b} \ket{b}, \\
	\ket{ba} = \frac{1}{\sqrt{N-1}} \sum_b \ket{b} \otimes \ket{a}, \\
	\ket{bb} = \frac{1}{\sqrt{N-1}} \sum_b \ket{b} \otimes \frac{1}{\sqrt{N+l-2}} \left( \sum_{b' \ne b} \ket{b'} + \sqrt{l} \ket{b} \right).
\end{eqnarray*}
In this basis, the initial state of the system is
\begin{eqnarray*}
	\ket{\psi_0} 
		= \frac{1}{\sqrt{N(N+l-1)}} \Big[ &\sqrt{l} \ket{aa} +\sqrt{N-1} \ket{ab} + \sqrt{N-1} \ket{ba} \\
		&+ \sqrt{(N-1)(N+l-2)} \ket{bb} \Big].
\end{eqnarray*}
The system evolves by repeatedly applying
\[ U' = SCQ, \]
where $Q$ is an oracle query that flips the amplitudes at the marked vertex, $C$ is the Grover diffusion coin as before \eref{eq:C}, and $S$ is the flip-flop shift. So $U'$ performs an oracle query followed by a step of the quantum walk. In the 4D subspace, it is
\begin{equation*}
	U' = \left( \!\! \begin{array}{cccc}
		\cos\theta & -\sin\theta & 0 & 0 \\
		0 & 0 & -\cos\phi & \sin\phi \\
		-\sin\theta & -\cos\theta & 0 & 0 \\
		0 & 0 & \sin\phi & \cos\phi \\
	\end{array} \!\! \right),
\end{equation*}
where
\[ \cos\theta = \frac{N-l-1}{N+l-1}, \quad {\rm and} \quad \sin\theta = \frac{2\sqrt{l(N-1)}}{N+l-1}, \]
and
\[ \cos\phi = \frac{N+l-3}{N+l-1}, \quad {\rm and} \quad \sin\phi = \frac{2\sqrt{N+l-2}}{N+l-1}. \]

\begin{figure}
\begin{center}
	\subfloat[]{
		\includegraphics{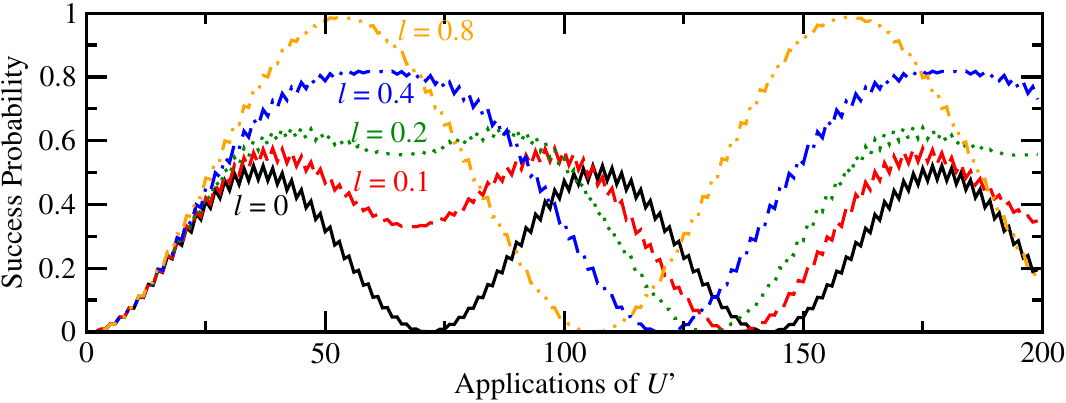}
		\label{fig:complete_N1024_less}
	} \quad
	\subfloat[]{
		\includegraphics{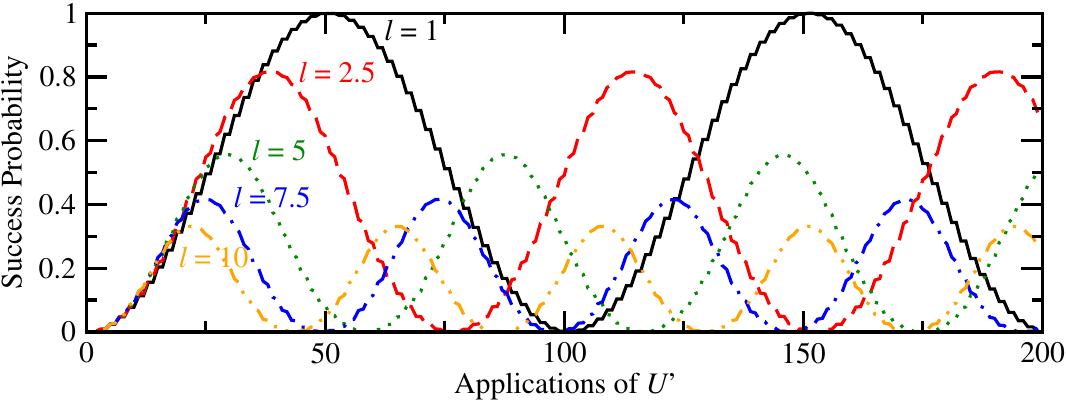}
		\label{fig:complete_N1024_greater}
	}
	\caption{\label{fig:complete_N1024} Success probability for search on the complete graph of $N = 1024$ vertices and a self-loop of weight $l$ at each vertex. (a) The solid black curve is $l = 0$, the dashed red curve is $l = 0.1$, the dotted green curve is $l = 0.2$, the dot-dashed blue curve is $l = 0.4$, and the dot-dot-dashed orange curve is $l = 0.8$. (b) The solid black curve is $l = 1$, the dashed red curve is $l = 2.5$, the dotted green curve is $l = 5$, the dot-dashed blue curve is $l = 7.5$, and the dot-dot-dashed orange curve is $l = 10$.}
\end{center}
\end{figure}

The probability of finding the particle at the marked vertex $a$ after $t$ steps is given by $p(t) = \left| \langle aa | U'^t | \psi_0 \rangle \right|^2 + \left| \langle ab | U'^t | \psi_0 \rangle \right|^2 $. This is plotted in \fref{fig:complete_N1024} as the system evolves for various values of $l$. In the loopless case, the success probability reaches $1/2$ after $\pi\sqrt{N}/2\sqrt{2}$ applications of $U'$. As $l$ increases, the maximum success probability increases until it reaches $1$ at $l = 1$. Further increasing $l$ causes the success probability to decrease. We also see in \fref{fig:complete_N1024_less} that when $l < 1/3$, the peak contains two humps, while when $l \ge 1/3$, the peak only has one hump. This transition at $l = 1/3$ will be proved rigorously, and it foreshadows that the analysis from \cite{Wong10}, while applying to $l \ge 1/3$, does not apply to $l < 1/3$.

Since $U'$ is a 4D matrix, it has four eigenvectors and corresponding eigenvalues, given in \cite{Wong10}. The initial state can be expressed in terms of these eigenvectors, and then it is straightforward to determine the state after $t$ applications of $U'$. For large $N$, it is
\[ U'^t \ket{\psi_0} \approx \left( \begin{array}{c}
	\frac{[1-\cos(\alpha t)] \sqrt{l(N-1)}}{(l+1)\sqrt{N+l-2}} \\
	\frac{2l + (l-1) \cos(\alpha t) + \sqrt{(2N+l-3)(l+1)} \sin(\alpha t)}{2(l+1)\sqrt{N+l-2}} \\
	\frac{2l + (l-1) \cos(\alpha t) - \sqrt{(2N+l-3)(l+1)} \sin(\alpha t)}{2(l+1)\sqrt{N+l-2}} \\
	\frac{1 + \cos(\alpha t)}{l + 1} \\
\end{array} \right), \]
where
\[ \cos\alpha = \frac{N-2}{N+l-1}, \quad {\rm and} \quad \sin\alpha = \frac{\sqrt{(2N+l-3)(l+1)}}{N+l-1}. \]
Then the success probability $p(t)$ is asymptotically given by the sum of the squares of the first two terms:
\begin{eqnarray*}
	p(t) &\approx \left[ \frac{[1-\cos(\alpha t)] \sqrt{l(N-1)}}{(l+1)\sqrt{N+l-2}} \right]^2 \\
	&\quad+ \left[ \frac{2l + (l-1) \cos(\alpha t) + \sqrt{(2N+l-3)(l+1)} \sin(\alpha t)}{2(l+1)\sqrt{N+l-2}} \right]^2.
\end{eqnarray*}
For large $N$, this further simplifies to
\begin{eqnarray*}
	p(t) &\approx \left[ \frac{[1-\cos(\alpha t)] \sqrt{l(N-1)}}{(l+1)\sqrt{N+l-2}} \right]^2 \\
	&\quad+ \left[ \frac{\sqrt{(2N+l-3)(l+1)} \sin(\alpha t)}{2(l+1)\sqrt{N+l-2}} \right]^2.
\end{eqnarray*}
To find the maximum success probability and the time at which it occurs, we take the first derivative of this, yielding
\[ \frac{dp}{dt} = \frac{\alpha \sin(\alpha t)}{2(l+1)^2(N+l-2)} \left[ 4l(N-1) + (2N-l-3)(1-l) \cos(\alpha t) \right]. \]
Setting this equal to zero, we get two solutions for the runtime $t_*$:
\[ t_* = \frac{\pi}{\alpha}, \quad t_* = \frac{1}{\alpha} \cos^{-1} \left( \frac{4l(N-1)}{(2N-l-3)(l-1)} \right). \]
The first solution $t_* = \pi/\alpha$ was explored in \cite{Wong10}. For the second solution, the argument of the inverse cosine is $2l/(l-1)$ for large $N$. This argument has magnitude less than $1$ when $l < 1/3$ and greater than $1$ when $l > 1/3$, resulting in real and complex runtimes, respectively. Thus we use this solution when $l < 1/3$ and the first solution $\pi/\alpha$ when $l \ge 1/3$. Approximating $\alpha \approx \sqrt{2(l+1)/N}$ for large $N$ and $l = o(N)$, we get a runtime of
\[ t_* \approx \frac{\cos^{-1} \left( \frac{2l}{l-1} \right)}{\sqrt{2(l+1)}} \sqrt{N} \]
when $l < 1/3$. Plugging this into $p(t)$ and keeping the dominant term for large $N$, the corresponding success probability is
\[ p_* \approx \frac{1}{2(1-l)}. \]

Putting these $l < 1/3$ results together with the $l \ge 1/3$ results from \cite{Wong10}, we get
\[ t_* \approx \left\{ \begin{array}{ll}
	{\cos^{-1} \left( \frac{2l}{l-1} \right) \over \sqrt{2(l+1)}} \sqrt{N} & l < 1/3 \\
	{\pi \over \sqrt{2(l+1)}} \sqrt{N} & l \ge 1/3,\ l = o(N) \\
	\pi / \sin^{-1} \left( {\sqrt{c(c+2)} \over {c+1}} \right) & l = cN \\
	2 & l = \omega(N) \\
\end{array} \right. \]
and
\[ p_* \approx \left\{ \begin{array}{ll}
	{1 \over 2(l+1)} & l < 1/3 \\
	{4l \over (l+1)^2} & l \ge 1/3,\ l = o(N) \\
	{16+9c \over 4c(c+1)} {1 \over N} & l = cN \\
	{9 \over 4l} & l = \omega(N) \\
\end{array} \right. . \]
Thus we have completely characterized the runtime and success probability of the algorithm for real $l \ge 0$.

Finally, note that the success probability is greater than the loopless value of $1/2$ when
\[ \frac{4l}{(l+1)^2} > \frac{1}{2} \quad \Rightarrow \quad l < 3 + 2\sqrt{2} \approx 5.828. \]
Recall that the standard lackadaisical quantum walk with integer $l$ self-loops per vertex had $l \le 5$, so the generalized walk has a larger range of values of $l$ that boost the success probability of a single iteration of the search algorithm.

%-------------------------------------------------------------------------------
% Section
%-------------------------------------------------------------------------------

\section{Conclusion}

Quantum walks are one of the primary means of developing quantum algorithms. The utility of walking on weighted graphs has been explored in several contexts, and we defined a discrete-time coined quantum walk on weighted graphs. The resulting coin operation is no longer an inversion about the average. With the flip-flop shift, two applications of this coined quantum walk is exactly equivalent to one application of Szegedy's quantum walk. With the moving shift, however, it is a new type of walk.

This allows lackadaisical quantum walks to be reduced and generalized by replacing, at each vertex, the $l$ identically-evolving self-loops with a single self-loop of weight $l$. When $l$ is an integer, the resulting walks are identical, but now $l$ can also take non-integer values.

We explored this for a walk on the line, showing that {\v{S}}tefa{\v{n}}{\'a}k \textit{et al}'s \cite{Stefanak2012} deformation of the three-state Grover walk is precisely the generalized lackadaisical quantum walk. This utilizes the moving shift, and the analytics of the evolution with the flip-flop shift is an open question.

We also explored search on the complete graph, which is equivalent to Grover's unstructured search problem. In doing so, the results for a regular lackadaisical quantum walk with $l$ self-loops per vertex directly carry over to the generalized case when $l \ge 1/3$. We provided new analysis, however, for when $l < 1/3$. This completely characterizes the generalized lackadaisical quantum walk for this search problem.

Further research includes applications of the generalized lackadaisical quantum walk to other quantum walk-based algorithms or spatial search problems.

%-------------------------------------------------------------------------------
% Acknowledgments.
%-------------------------------------------------------------------------------

\ack
This work was supported by the U.S.~Department of Defense Vannevar Bush Faculty Fellowship of Scott Aaronson.

%-------------------------------------------------------------------------------
% References.
%-------------------------------------------------------------------------------

\section*{References}
\bibliographystyle{iop}
\bibliography{refs}

\begin{thebibliography}{10}
\providecommand{\url}[1]{\texttt{#1}}
\providecommand{\urlprefix}{URL }
\providecommand{\eprint}[2][]{\url{#2}}

\bibitem{Childs2009}
Childs A~M 2009 Universal computation by quantum walk \emph{Phys. Rev. Lett.}
  \textbf{102} 180501

\bibitem{SKW2003}
Shenvi N, Kempe J and Whaley K~B 2003 Quantum random-walk search algorithm
  \emph{Phys. Rev. A} \textbf{67} 052307

\bibitem{Ambainis2004}
Ambainis A 2004 Quantum walk algorithm for element distinctness \emph{Proc.
  45th Annual IEEE Symp. Found. Comput. Sci.} FOCS '04 (IEEE Computer Society)
  pp. 22--31

\bibitem{FGG2008}
Farhi E, Goldstone J and Gutmann S 2008 A quantum algorithm for the
  {H}amiltonian {NAND} tree \emph{Theory Comput.} \textbf{4}(8) 169--190

\bibitem{Christandl2004}
Christandl M, Datta N, Ekert A and Landahl A~J 2004 Perfect state transfer in
  quantum spin networks \emph{Phys. Rev. Lett.} \textbf{92} 187902

\bibitem{Kendon2011}
Kendon V~M and Tamon C 2011 Perfect state transfer in quantum walks on graphs
  \emph{Journal of Computational and Theoretical Nanoscience} \textbf{8}(3)
  422--433

\bibitem{Carlson2007}
Carlson W, Ford A, Harris E, Rosen J, Tamon C and Wrobel K 2007 Universal
  mixing of quantum walk on graphs \emph{Quantum Inf. Comput.} \textbf{7}(8)
  738--751

\bibitem{Wong16}
Wong T~G 2015 Faster quantum walk search on a weighted graph \emph{Phys. Rev.
  A} \textbf{92} 032320

\bibitem{Wong22}
Wong T~G and Philipp P 2016 Engineering the success of quantum walk search
  using weighted graphs \emph{Phys. Rev. A} \textbf{94} 022304

\bibitem{Szegedy2004}
Szegedy M 2004 Quantum speed-up of {M}arkov chain based algorithms \emph{Proc.
  45th Annual IEEE Symp. Found. Comput. Sci.} FOCS '04 (Washington, DC, USA:
  IEEE Computer Society) pp. 32--41

\bibitem{Magniez2012}
Magniez F, Nayak A, Richter P~C and Santha M 2012 On the hitting times of
  quantum versus random walks \emph{Algorithmica} \textbf{63}(1) 91--116

\bibitem{Wong26}
Wong T~G 2017 Equivalence of {S}zegedy's and coined quantum walks \emph{Quantum
  Inf. Process.} \textbf{16}(9) 215

\bibitem{Wong10}
Wong T~G 2015 Grover search with lackadaisical quantum walks \emph{J. Phys. A:
  Math. Theor.} \textbf{48}(43) 435304

\bibitem{Wang2016}
Wang K, Wu N, Xu P and Song F 2016 The lackadaisical quantum walker is {NOT}
  lazy at all \emph{{a}rXiv:1612.03370 [quant-ph]}

\bibitem{Stefanak2012}
{\v{S}}tefa{\v{n}}{\'a}k M, Bezd{\v{e}}kov{\'a} I and Jex I 2012 Continuous
  deformations of the grover walk preserving localization \emph{The European
  Physical Journal D} \textbf{66}(5) 142

\bibitem{Grover1996}
Grover L~K 1996 A fast quantum mechanical algorithm for database search
  \emph{Proc. 28th Annual ACM Symp. Theory Comput.} STOC '96 (New York, NY,
  USA: ACM) pp. 212--219

\bibitem{Meyer1996a}
Meyer D~A 1996 From quantum cellular automata to quantum lattice gases \emph{J.
  Stat. Phys.} \textbf{85}(5-6) 551--574

\bibitem{Aharonov2001}
Aharonov D, Ambainis A, Kempe J and Vazirani U 2001 Quantum walks on graphs
  \emph{Proc. 33rd Annual ACM Symp. Theory Comput.} STOC '01 (New York, NY,
  USA: ACM) pp. 50--59 ISBN 1-58113-349-9

\bibitem{Moore2002}
Moore C and Russell A 2002 Quantum walks on the hypercube \emph{Proc. 6th Int.
  Workshop on Randomization and Approximation Techniques in Comput. Sci.}
  RANDOM 2002 (Berlin, Heidelberg: Springer) pp. 164--178 ISBN
  978-3-540-45726-8

\bibitem{Ambainis2001}
Ambainis A, Bach E, Nayak A, Vishwanath A and Watrous J 2001 One-dimensional
  quantum walks \emph{Proc. 33rd Annual ACM Symp. Theory Comput.} STOC '01 (New
  York, NY, USA: ACM) pp. 37--49 ISBN 1-58113-349-9

\bibitem{AKR2005}
Ambainis A, Kempe J and Rivosh A 2005 Coins make quantum walks faster
  \emph{Proc. 16th Annual ACM-SIAM Symp. Discrete Algorithms} SODA '05
  (Philadelphia, PA, USA: SIAM) pp. 1099--1108 ISBN 0-89871-585-7

\bibitem{Wong23}
Ambainis A, Pr\={u}sis K, Vihrovs J and Wong T~G 2016 Oscillatory localization
  of quantum walks analyzed by classical electric circuits \emph{Phys. Rev. A}
  \textbf{94} 062324

\bibitem{Inui2005}
Inui N, Konno N and Segawa E 2005 One-dimensional three-state quantum walk
  \emph{Phys. Rev. E} \textbf{72} 056112

\bibitem{Wong17}
Wong T~G 2015 Quantum walk on the line through potential barriers \emph{Quantum
  Inf. Process.} \textbf{15}(2) 675--688

\end{thebibliography}

\end{document}